%% file: QOCE.tex
\documentclass[twoside,leqno,twocolumn]{article}
\usepackage[letterpaper]{geometry}
\usepackage{ltexpprt}
\usepackage{cite}
\usepackage{amsmath,amssymb,amsfonts}
\usepackage{graphicx}
\usepackage{subfigure}
\usepackage{textcomp}
\usepackage{xcolor}
\usepackage{hyperref}
\usepackage{threeparttable}
\usepackage{algorithm}
\usepackage{algorithmic}

\begin{document}
\newcommand{\KunHe}[1]{\textcolor{red}{[KunHe: #1]}}
\newcommand{\Yang}[1]{\textcolor{blue}{[Yang: #1]}}

\title{\Large Quadratic Optimization based Clique Expansion \\for Overlapping Community Detection  
}

\author{Yanhao Yang \thanks{School of Computer Science and Technology, Huazhong University of Science and Technology, Wuhan, China.}
\and Pan Shi $^\ast$
\and Yuyi Wang \thanks{Department Electrical Engineering, ETH, Zürich, Switzerland.}
\and Kun He $^\ast$\thanks{Corresponding author, email: brooklet60@hust.edu.cn.}
}


\date{}

\maketitle


\begin{abstract}
Community detection is crucial for analyzing social and biological networks, and comprehensive approaches have been proposed 
in the last two decades. Nevertheless, finding all overlapping communities 
in large networks that could accurately approximate the ground-truth communities remains challenging. 
In this work, we present the QOCE (Quadratic Optimization based Clique Expansion), an overlapping community detection algorithm that could scale to large networks with hundreds of thousands of nodes and millions of edges. QOCE follows the popular seed set expansion strategy, regarding each high-quality maximal clique as the initial seed set and applying quadratic optimization for the expansion. 
We extensively evaluate our algorithm on 28 synthetic LFR networks and six real-world networks of various domains and scales, and compare QOCE with four state-of-the-art overlapping community detection algorithms. Empirical results demonstrate the competitive performance of the proposed approach in terms of detection accuracy, efficiency, and scalability.
\end{abstract}

\section{Introduction}
\label{sec:Introduction}
\input{tex/Introduction}

\section{Related Work}
\label{sec:RelatedWork}
\input{tex/RelatedWork}

\section{The Proposed QOCE Method}
\label{sec:ProposedMethod}
\input{tex/ProposedMethod}

\section{Experimental Results}
\label{sec:Experiments}
\input{tex/Experiments}

\section{Conclusion}
\label{sec:Conclusion}
\input{tex/Conclusion}


\bibliographystyle{IEEEtran}
\bibliography{QOCE}

\end{document}

%% file: tex/Introduction.tex
In network science, a community is usually defined as a group of nodes whose internal connections are much denser than their external connections to the remainder of the network. Community detection in networks plays a crucial role in various applications, including functional prediction in biology, online social network analysis, and anomaly detection in networks, etc.

Early community detection research focuses on graph partitioning \cite{fortunato2010community}, which is often based on the modularity optimization or spectral clustering \cite{newman2006finding,von2007tutorial}. Their outputs are disjoint communities, contradicting that a single node may have multiple community memberships in real-world networks. For instance, a user in a social network may have several social circles (e.g., ``lists" on Facebook or Twitter), and then the user is in the overlapping area of the social circles. Thus, overlapping community detection has drawn much more attention due to its reasonable assumption~\cite{xie2013overlapping}. 
\begin{figure}
\centering
\includegraphics[scale=0.28]{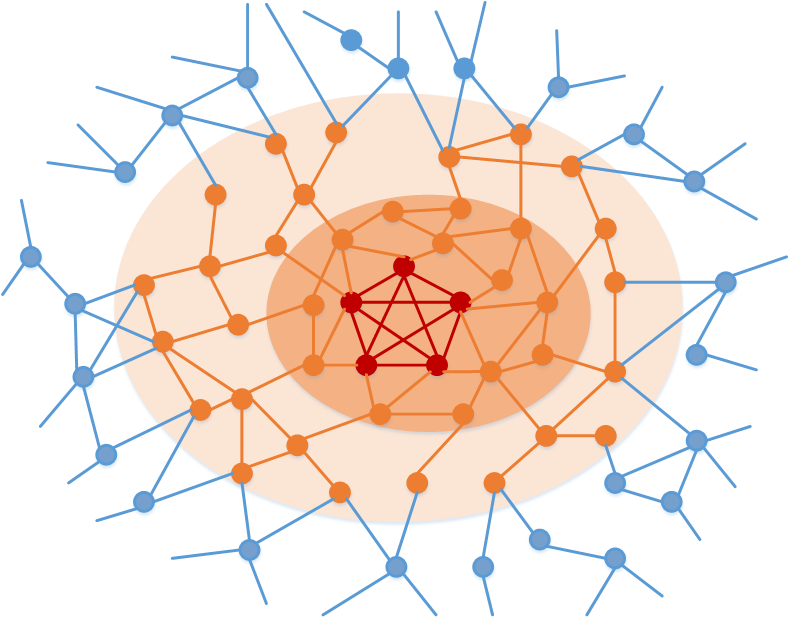}
\vspace{-0.5em}
\caption{An illustration of our clique expansion method. The inner clique with five vertices (in red) is the clique seed set, then we sample a subgraph (in light orange) around the seed set and extract a community (in dark orange) inside the subgraph.}
\label{Fig:expansion}
\vspace{-1.3em}
\end{figure}

In this work, we present a new overlapping community detection approach, termed Quadratic Optimization based Clique Expansion (QOCE), which falls into the category of seed set expansion, in particular, the category of clique expansion. Seed Set Expansion (SSE) is a typical class of methods for overlapping community detection. 
Given some sets of nodes, called the \textit{seeds} or \textit{seed sets}, the SSE methods expand from each seed set to obtain a community, and the expansion is achieved by iteratively adding proper nodes from the neighborhood into the community. From different seed sets, we can get different and possibly overlapping communities. Note that this method does not require that every node has community memberships, consistent with the situation on real-world networks. 

On the one hand, for SSE-based community detection methods, the quality of the detected communities depends on the seeding strategy. 
For the initial seed set, we notice that \textit{cliques} have the densest internal connections, and the community usually contains one or more cliques, so cliques could be natural cores for community structures. For instance, Clique Percolation \cite{palla2005uncovering} detects communities by merging $k$-cliques ($k$ is a parameter) that share $k-1$ nodes. Furthermore, nodes in a maximal clique generally belong to the same community. 
Due to the properties of maximal cliques, it makes sense to regard each maximal clique with at least $k$ nodes as the initial seed set for community detection \cite{shen2009detect,lee2010detecting}.

On the other hand, the detection quality also depends on the expansion method. Previous methods usually expand the seed sets globally in a greedy way \cite{palla2005uncovering,whang2016overlapping}. We consider splitting the expansion phase into two parts: sampling and extraction, as illustrated in Figure \ref{Fig:expansion}. We first adopt a fast diffusion method from each seed set to sample a subgraph of proper size quickly and then extract a community from the sampled subgraph.
To quickly sample the subgraph, we implement a lazy random walk with no matrix operation. 
For extracting a community from the sampled subgraph, we use Cheeger cut \cite{chung1997spectral} as the criteria for determining the community boundary. The objective of this extraction is to minimize the Cheeger cut of the target community. As it is hard to solve the Cheeger cut minimization problem directly, we approximately transform the problem into a quadratic optimization problem for community extraction. 
Finally, a post-processing is added to accurately determine the ultimate community memberships by a local minimum of the conductance. 
Furthermore, as each clique expansion is independent, we also implement a parallel version of QOCE.

Our main contributions are in three-fold:
\vspace{-1.2em}
\begin{itemize}
\setlength{\itemsep}{0pt}
\setlength{\parsep}{0pt}
\setlength{\parskip}{0pt}
	\item We propose a new algorithm for overlapping community detection, which outperforms state-of-the-art overlapping community detection approaches in terms of the detection accuracy in real-world and synthetic networks, and the proposed algorithm can scale to large networks.
	\item Regarding each high-quality maximal clique as the initial seed set, we adopt a fast random walk diffusion method to sample a comparatively small subgraph from the clique quickly. This sampling strategy differs from existing seed set expansion methods for global community detection and greatly reduces the subsequent computational overhead.
	\item We introduce quadratic optimization for global community detection. Motivated by Cheeger cut \cite{chung1997spectral}, we design a quadratic objective function with $l_1$ regularization term on the sampled subgraph and obtain a sparse affiliation vector by solving the quadratic optimization problem.
\end{itemize}


%% file: tex/RelatedWork.tex
In the past two decades, various overlapping community detection methods have been proposed, including link partition \cite{evans2009line}, label propagation \cite{coscia2012demon}, NMF-based methods \cite{yang2013overlapping,ye2019discrete}, and other 
\cite{gujral2018smacd}. 
Among which, Seed Set Expansion (SSE)~\cite{palla2005uncovering,shen2009detect,lee2010detecting,lancichinetti2011finding,whang2016overlapping} is a typical approach. 
In the following, we will discuss two aspects related to our method: various seed set expansion approaches and optimization methods used for community detection.

\textbf{Seed Set Expansion.}
One typical stream of overlapping community detection is seed set expansion. Many of these methods choose maximal cliques as the seed sets. For instance, Clique Percolation \cite{palla2005uncovering}, the first proposed algorithm for overlapping community detection, starts from each maximal $k$-clique and merges cliques sharing $k-1$ nodes to form a percolation chain. 
Similarly, Shen \emph{et al.} \cite{shen2009detect} and Lee \emph{et al.} \cite{lee2010detecting} choose maximal cliques as the seeds, 
but Shen \emph{et al.} \cite{shen2009detect} adopts an agglomerative clustering framework to uncover both the overlapping and hierarchical community structure in complex networks, while Lee \emph{et al.} \cite{lee2010detecting} expands these seeds by greedily optimizing a local fitness function. 

There are some seed set expansion methods that adopt other seeding strategies. Lancichinetti \emph{et al.} \cite{lancichinetti2011finding} treat randomly picked node as the initial seed and optimizes a fitness function, defined as the probability of finding the community in a random null model, to join together small communities into statistically significant larger communities. Whang \emph{et al.} \cite{whang2016overlapping} try four different seeding strategies and expand seeds by personalized PageRank diffusion. 

The seed sets of our proposed QOCE are also composed of maximal cliques. Inspired by the local community detection methods~\cite{HeSBH/TKDD2019} that usually sample a subgraph for a given seed set, QOCE adopts a fast random walk for the expansion to sample a comparatively small subgraph from each clique seed set quickly, thereby significantly reduces the subsequent calculation.

\textbf{Optimizations for Community Detection.}
Various optimization approaches have been used for community detection, such as conductance minimization \cite{whang2016overlapping} and likelihood maximization \cite{yang2013overlapping}. Researchers usually design a greedy algorithm to approximate the optimization problem \cite{whang2016overlapping}, or treat the optimization problem as a regular convex optimization problem to be solved by gradient methods \cite{yang2013overlapping}.

Since quadratic optimization has more requirements, its application is relatively rare for community detection. Zhang et al.\ \cite{zhang2013robust} use a half-quadratic technique to solve the objective optimization problem with correntropy as a robust regularization term for subspace clustering. Su and Havens \cite{su2015quadratic} propose several highly efficient and effective heuristics based on quadratic optimization to address the fuzzy modularity detection.

%% file: tex/ProposedMethod.tex
Given an undirected graph $G=(\mathcal{V},\mathcal{E})$ with vertex set $\mathcal{V}$ and edge set $\mathcal{E}$, which can be presented as a symmetric adjacent matrix $\mathbf{A} \in \{0,1\}^{|\mathcal{V}| \times |\mathcal{V}|}$, 
the goal of overlapping community detection is to mine all overlapping communities $\mathcal{C} = \{C_1, C_2, ..., C_{|\mathcal{C}|}\}$. Each community $C \in \mathcal{C}$ is a subset of vertex set $\mathcal{V}$, and the union of these communities does not necessarily cover the entire vertex set $\mathcal{V}$. These communities have comparatively denser internal links and comparatively sparser external links in graph $G$.

There are four phases in our proposed Quadratic Optimization based Clique Expansion (QOCE) algorithm, namely seeding, sampling, extraction, and post-processing.

\subsection{Seeding.}
In this phase, we first enumerate all maximal cliques in graph $G$, and then select high-quality maximal cliques to form the initial seed sets $\mathcal{S}$. A \textit{clique} is a complete subgraph such that every two distinct vertices are connected. A \textit{maximal clique} is a clique that is not contained in any larger clique.

In general, enumerating all maximal cliques in a graph is computationally expensive, however, maximal cliques can be found quickly in graphs that are sufficiently sparse. To this end, our implementation apply the Bron-Kerbosch (BK) algorithm \cite{bron1973algorithm}, a recursive backtracking algorithm, to efficiently find all maximal cliques with at least four vertices. To speedup the computation, before applying the BK algorithm, we introduce a preprocessing procedure to recursively remove vertices with degree less than three to obtain $k-$\textit{cores} ($k = 3$) in the graph. For these $k-$cores, we run the BK algorithm sequentially or in parallel.

For ``high-quality"  maximal cliques, we mean they are large in sizes and do not overlap with each other heavily. If highly overlapping maximal cliques are all treated as seeds, we are likely to obtain almost duplicate communities. Therefore, we employ a simple filtering method to exclude highly overlapping maximal cliques. In particular, we first sort all maximal cliques by sizes in the non-ascending order, and then remove subsequent cliques with at least $75\%$ (which is an appropriate threshold for seed exclusion, and also adopted by \cite{lee2010detecting}) of vertices contained in any previous clique.

\subsection{Sampling.}

In large networks, performing seed set expansion in the entire graph is computationally expensive. Inspired by the work of local community detection~\cite{HeSBH/TKDD2019}, which only cares about finding the local community of a given seed set and does local sampling to reduce the subsequent computational overhead, we do a local sampling from each clique seed set to reduce the subsequent calculation. Due to the ``local" property of a community, that is, a vertex will not go too far in reaching other vertices in its community, as the \textit{diameter} of the community (i.e., the maximum pairwise distance among vertices) are usually small. 
To sample a subgraph covering the community, we could do short random walks from each clique seed set. 

Given a clique seed set $S \in \mathcal{S}$, for which we hope to find a community, the initial probability distribution vector $\mathbf{p}_0$ is defined by:
\begin{equation}
\mathbf{p}_0(v) = 
\begin{cases}
	\mathbf{d}(v)/\text{Vol}(S) & \text{if} ~v\in S, \\
	0 & \text{otherwise}.
\end{cases}
\label{Eq:init_prob}
\end{equation}
Here $\mathbf{d}(v)$ is the degree of vertex $v$ and $\text{Vol}(S)$ is the total degree of the vertices in $S$.

In order to retain some probability at the current vertex, we adopt lazy random walk, and the transition matrix is as follows:
\begin{equation}
\mathbf{N_{rw}} = (\mathbf{D} + \mathbf{I})^{-1} (\mathbf{A} + \mathbf{I}),
\label{Eq:tran_matrix}
\end{equation}
where the degree matrix $\mathbf{D} \in \mathbb{Z}^{|\mathcal{V}| \times |\mathcal{V}|}$ is defined as a diagonal matrix whose $i\text{-th}$ diagonal entry is $\sum_j a_{ij}$, and the identity matrix $\mathbf{I}$ is of order $|\mathcal{V}|$.
So we can calculate the probability distribution vector of $t$-step lazy random walks by:
\begin{equation}
\mathbf{p}_{t} = (\mathbf{N_{rw}} ^ T)^{t} \mathbf{p}_0.
\label{Eq:new_prob}
\end{equation}

It is a typical approach to obtain the probability distribution $\mathbf{p}_{t}$ by (\ref{Eq:new_prob}). However, it is unpractical to compute on the transition matrix $\mathbf{N_{rw}}$ for a large graph. We implement a fast method to gain $\mathbf{p}_{t}$ and the corresponding sampling graph. See details in Algorithm \ref{alg:random_walk}, where $N(s_i)$ denotes the neighbourhood (set of adjacent vertices) of vertex $s_i$ in $G$.

\begin{algorithm}
\begin{algorithmic}[1]
\REQUIRE Graph $G = (\mathcal{V},\mathcal{E})$, clique seed set $S \in \mathcal{S}$, number of steps $t_0$, and threshold $\mu$
\ENSURE Sampled subgraph $G_s = (\mathcal{V}_s, \mathcal{E}_s)$
\STATE  Initialize $t = 0$, $\mathbf{p}_0$ by (\ref{Eq:init_prob})
\WHILE {($t < t_0$)}
\STATE $U = \emptyset$
\STATE $\mathbf{p}_{t+1} = \mathbf{0}$
\FOR{each $s_i \in S$}
\STATE $U_i = N(s_i)$ 
\STATE $\mathbf{p}_{t+1}(s_i) = \mathbf{p}_{t+1}(s_i) + \mathbf{p}_t(s_i)/(|U_i|+1)$
\FOR {each $v_i \in U_i$}
\STATE $\mathbf{p}_{t+1}(v_i) = \mathbf{p}_{t+1}(v_i) + \mathbf{p}_t(s_i)/(|U_i|+1)$
\ENDFOR
\STATE $U = U \cup U_i \cup {s_i}$
\ENDFOR
\STATE $S = U$
\STATE $t = t + 1$
\ENDWHILE
\STATE $\mathcal{V}_s \leftarrow$ \text{vertices whose entry value $> \mu$ in} $\mathbf{p}_{t_0}$
\STATE $G_s = (\mathcal{V}_s,\mathcal{E}_s)$ is the induced subgraph from $\mathcal{V}_s$
\caption{Fast random walk sampling}
\label{alg:random_walk}
\end{algorithmic}
\end{algorithm}

For each clique seed set $S$ in seed sets $\mathcal{S}$, we sample a comparatively small subgraph $G_s$ by Algorithm \ref{alg:random_walk}. This procedure can be completed in milliseconds for large networks with millions of vertices. All subsequent procedures are then carried on subgraph $G_s$, rather than on the entire graph $G$, which significantly reduces the computational overhead.

\subsection{Extraction.}
To extract a community $C$ with most sparse external links from $G_s=(\mathcal{V}_s, \mathcal{E}_s)$, we employ the Cheeger cut \cite{chung1997spectral} and derive the following optimization problem:
\begin{equation}
\min_{C \subset \mathcal{V}_s} \psi(C) = \frac{\text{cut}(C,\overline{C})}{\text{min}\{|C|,|\overline{C}|\}},
\label{Eq:ori_opt}
\end{equation}
where vertex set $\overline{C} = \mathcal{V}_s / C$, and $\text{cut}(C, \overline{C})$ denotes the number of edges between $C$ and $\overline{C}$.

Let $\mathbf{L}_s = \mathbf{D}_s - \mathbf{A}_s$ be the Laplacian matrix of $G_s$ where $\mathbf{A}_s$ and $\mathbf{D}_s$ are the adjacency matrix and the diagonal degree matrix of $G_s$, respectively. 
Let $\mathbf{y} \in \{0,1\}^{|\mathcal{V}_s|}$ be an affiliation vector with each entry $\mathbf{y}_i$ representing whether vertex $v_i$ belongs to the detected community $C$. Then, $\mathbf{y}^T \mathbf{y}$, $\mathbf{y}^T \mathbf{D}_s \mathbf{y}$, and $\mathbf{y}^T \mathbf{A}_s \mathbf{y}$ can be respectively treated as the number of vertices, the total degree of vertices, and two times the number of internal edges in $C$~\cite{HeSBH/TKDD2019}. So $\text{cut}(C,\overline{C}) = \mathbf{y}^T \mathbf{D}_s \mathbf{y} - \mathbf{y}^T \mathbf{A}_s \mathbf{y} = \mathbf{y}^T \mathbf{L}_s \mathbf{y}$.

Community $C$ is relatively small in the sampled graph $G_s$, which means $|C| < \frac{1}{2}|\mathcal{V}_s|$. Then the Cheeger cut $\psi(C)$ could be written as a Rayleigh quotient, and the problem in (\ref{Eq:ori_opt}) could be rewritten as follows:
\begin{equation}
\begin{aligned}
\min_{\mathbf{y}} ~&~ \frac{\mathbf{y}^T \mathbf{L}_s \mathbf{y}}{\mathbf{y}^T \mathbf{y}} \\
\emph{s.t.} ~&~ y_i \in \{0,1\} ~~ \forall v_i \in \mathcal{V}_s,
\end{aligned}
\label{Eq:int_opt}
\end{equation}
where $y_i$ denotes the $i\text{-th}$ entry of the affiliation vector $\mathbf{y}$, and $v_i$ denotes the $i\text{-th}$ vertex in $G_s$.

Due to the constraint on $\mathbf{y}$, (\ref{Eq:int_opt}) is a 0-1 integer programming problem that is NP-complete. We relax $\mathbf{y}$ from $\{0,1\}^{|\mathcal{V}_s|}$ to  $[0,1]^{|\mathcal{V}_s|}$, and note that $\mathbf{y}^T \mathbf{y} = \mathbf{e}^T \mathbf{y}$ where vector $\mathbf{e}$ is $\mathbf{1}$. Thus, we relax the 0-1 integer programming problem to the following quadratic optimization problem:
\begin{equation}
\begin{aligned}
\min_{\mathbf{y}} ~&~ \mathbf{y}^T \mathbf{L}_s \mathbf{y} \\
\emph{s.t.} ~&~ \mathbf{e}^T \mathbf{y} = 1, \\
~&~ y_i \in [0,1] ~~ \forall v_i \in \mathcal{V}_s.
\end{aligned}
\label{Eq:quad_opt}
\end{equation}

There is a trick that the first constraint in (\ref{Eq:quad_opt}) can be regarded as an regulation term in the objective function. Expecting the detected community $C$ to contain the seed set $S$, we also add a new constraint to this optimization problem. Finally, the quadratic optimization problem in (\ref{Eq:quad_opt}) can be rewritten as follows: 
\begin{equation}
\begin{aligned}
\min_{\mathbf{y}} ~&~ \mathbf{y}^T \mathbf{L}_s \mathbf{y} + \alpha \mathbf{e}^T \mathbf{y} \\
\emph{s.t.} ~&~ y_i \in [0,1] ~~ \forall v_i \in \mathcal{V}_s, \\
~&~ y_i \geq \frac{1}{|S|} ~~ \forall v_i \in S,
\end{aligned}
\label{Eq:final_opt}
\end{equation}
where parameter $\alpha$ is added to balance the two terms.

The quadratic programming problem in (\ref{Eq:final_opt}) can be efficiently solved by interior point method~\cite{altman1999regularized}. The optimal solution $\mathbf{y}^*$ is the affiliation between the community $C$ and all vertices in $G_s$.

\subsection{Post-processing.} 
The resulting $\mathbf{y}^*$ is a soft community membership with each entry $y^*_i$ representing the probability of vertex $v_i$ belonging to community $C$. In order to obtain the ultimate community membership, we need to post-process $\mathbf{y}^*$. Generally, the post-processing is to specify a threshold on $\mathbf{y}^*$ manually, but it is not a desirable approach \cite{ye2019discrete}. We introduce the conductance metric to determine the community boundary.

\begin{Definition}[Conductance]
For a subgraph $C$ whose vertex set $V_C \subset V_s$, its conductance \cite{kannan2004clusterings} is:
\begin{equation} \nonumber
\label{Eq:cond}
\Phi(C) = \frac{\text{cut}(C,\overline{C})}{\text{min}\{\text{Vol}(C),\text{Vol}(\overline{C})\}},
\end{equation}
where $\text{Vol}(C)$ denotes the total degree of vertices inside subgraph $C$.
\label{Def:cond}
\end{Definition}

We first sort the vertices based on the entry values of $\mathbf{y}^*$ in non-ascending order. Suppose the sorted sequence is $\{ v_1, v_2, ..., v_{|\mathcal{V}_s|} \}$. For each integer $k \in \{1,..., |\mathcal{V}_s|\}$, we extract a vertex set from $G_S$ with the first $k$ vertices in the sorted sequence, denoted by $\hat{C}_k = \{v_i|i<=k\}$. Then for each vertex set $\hat{C}_k$, we compute its conductance $\Phi(\hat{C}_k)$, and a vertex set $\hat{C}_{k^*}$ with the first local minimum conductance is the desired community $C$.

The method of finding a local minimal conductance of the community is shown in Algorithm \ref{alg:post_proc}. 
Empirical observation shows that not only the size of the community obtained by a local minimum is closer to the size of the ground-truth community than that obtained by the global minimum, but also a lot of unnecessary computational overhead is saved compared with the global minimum. Furthermore, we introduce a window $w$ to ensure $\Phi(\hat{C}_{k^*})$ is the minimum in $\{ \Phi(\hat{C}_1), \Phi(\hat{C}_2),...,\Phi(\hat{C}_{k^*}),\Phi(\hat{C}_{k^*+1}),...,\Phi(\hat{C}_{k^*+w}) \}$, which avoids short-term fluctuation of conductance.

\begin{algorithm}
\begin{algorithmic}[1]
\REQUIRE Sampled graph $G_s=(\mathcal{V}_s, \mathcal{E}_s)$, affiliation vector $\mathbf{y}^*$, window size $w$
\ENSURE Detected community $C$
\STATE Sort vertices in $\mathcal{V}_s$ by the corresponding values in $\mathbf{y}^*$ in non-ascending order, the sorted sequence is $\{ v_1, v_2, ..., v_{|\mathcal{V}_s|} \}$
\FOR {$k = 1:w$}
\STATE Calculate $\Phi(\hat{C}_k)$ by Definition \ref{Def:cond}
\ENDFOR
\STATE $k = 0, n = 0$
\WHILE {$n < w$}
\STATE $k = k + 1$
\STATE Calculate $\Phi(\hat{C}_{k+w})$ by Definition \ref{Def:cond}
\STATE $n = |\{\hat{C}_i|\Phi(\hat{C}_i)>\Phi(\hat{C}_k), \forall i, k < i \leq k+w\}|$
\ENDWHILE
\STATE $C = \{v_j | 1 \leq j \leq k\}$
\caption{Post-processing}
\label{alg:post_proc}
\end{algorithmic}
\end{algorithm}

Based on the above four phases, the overall Quadratic Optimization based Clique Expansion (QOCE) algorithm is shown in Algorithm \ref{alg:qoce}. 

\begin{algorithm}
\begin{algorithmic}[1]
\REQUIRE $G=(\mathcal{V}, \mathcal{E})$, 
         random walk steps $t_0$,
         random walk threshold $\mu$
         trade-off parameter $\alpha$, 
         and window size for post-processing $w$
\ENSURE A set of detected communities $\mathcal{C}$
\STATE Get $k-$cores ($k = 3$) in $G$
\STATE Find all maximal cliques $\mathcal{\hat{S}}$ by BK algorithm \cite{bron1973algorithm}
\STATE Get seed sets $\mathcal{S}$ by excluding highly overlapping cliques from $\mathcal{\hat{S}}$
\STATE $\mathcal{C} = \emptyset$
\FOR {each clique seed set $S \in \mathcal{S}$}
\STATE Sample $G_s = (\mathcal{V}_s,\mathcal{E}_s)$ from $S$ by Algorithm~\ref{alg:random_walk}
\STATE Obtain affiliation vector $\mathbf{y}^*$ by solving (\ref{Eq:final_opt})
\STATE Detect a community $C$ by Algorithm \ref{alg:post_proc}
\STATE $\mathcal{C} = \mathcal{C} \cup \{C\}$
\ENDFOR
\caption{The QOCE algorithm}
\label{alg:qoce}
\end{algorithmic}
\end{algorithm} 

\subsection{Time Complexity Analysis.}
For the maximal clique enumeration, we adopt the BK algorithm \cite{bron1973algorithm} with complexity $O(d|\mathcal{V}|3^{d/3})$, where $d$ is the degeneracy of the graph ($d<<|\mathcal{V}|$ in sparse networks).
The preprocessing of removing edges to get $k-$cores ($k = 3$) is $O(|\mathcal{V}|)$.
And maximal cliques filtering is $O(|\mathcal{S}|^2)$. The complexity of fast random walk, scaled with the output size of each sampled subgraph, is $|\mathcal{V}_s|$. So the total complexity of the sampling is $O(\sum_{S\in\mathcal{S}} |\mathcal{V}_s|)$. 

In the extraction phase, we employ interior point method \cite{altman1999regularized} to solve the quadratic optimization problem, in which the variable is an affiliation vector $\mathbf{y} \in [0,1]^{|\mathcal{V}_s|}$. And the interior point method requires $O(nL)$ iterations, where $n$ is the dimension of variable $\mathbf{y}$ in (\ref{Eq:final_opt}), that is $n=|\mathcal{V}_s|$, and $L$ is the length of a binary coding of the input data in problem (\ref{Eq:final_opt}), that is $L=\sum_{i=0}^{|\mathcal{V}_s|} \sum_{j=0}^{\mathcal{V}_s} (\lceil \log_2 (|L_{ij}| + 1) + 1 \rceil) + |S| \lceil \log_2 (\frac{1}{|S|} + 1) + 1 \rceil + (|\mathcal{V}_s| - |S|) \lceil \log_2 (0 + 1) + 1 \rceil$, in which $L_{ij}$ is the entry of the Laplacian matrix $\mathbf{L}_s$ and its value is usually equal to 0 in sparse networks. Thus, $L= |\mathcal{V}_s|^2 + |S| + |\mathcal{V}_s|$ for each subgraph, and the complexity of each extraction is $O(|\mathcal{V}_s|^3)$. The total complexity of this phase is $O(\sum_{S \in \mathcal{S}} |\mathcal{V}_s|^3)$.

Finally, our post-processing first sorts entries of the affiliation vector with complexity $O(|\mathcal{V}_s|log|\mathcal{V}_s|)$, then searches for the top $k$ vertices with local minimum conductance which is output-sensitive procedure with complexity $O(|C|)$. Note that the community is extracted from the sampled subgraph, $C \in \mathcal{V}_s$, the total complexity of this phase is $O(\sum_{S\in\mathcal{S}} |\mathcal{V}_s|log|\mathcal{V}_s|)$.

Hence, the overall complexity is $O(d|\mathcal{V}|3^{d/3} + |\mathcal{S}|^2 + \sum_{S\in\mathcal{S}} |\mathcal{V}_s|^3)$. 

\subsection{Parallel QOCE.} \label{sec:parallel}
The QOCE algorithm is a pipeline architecture consisting of four phases: seeding, sampling, extraction, and post-processing. The last three phases are a series of independent operations on each clique seed set. As shown in Algorithm \ref{alg:qoce}, line $5 \sim 8$ is a loop body, and each loop is independent. So we can run these loops in parallel without dependency issues.

Therefore, we design a parallel version of QOCE. We run the remaining phases on $h$ identical machines or CPUs in parallel after the seeding phase. Specifically, we regard each seed set's remaining phases as an individual task and each machine or CPU as an individual worker. We iteratively assign a new task to a free worker until all tasks have been executed. In this way, we reduce the running time of the last three phases to almost $1/h$ of the sequential version. The complexity of parallel QOCE is $O(d|\mathcal{V}|3^{d/3} + |\mathcal{S}|^2 + \frac{1}{h} \sum_{S\in\mathcal{S}} |\mathcal{V}_s|^3)$. 

Furthermore, after preprocessing, if the $k-$core ($k = 3$) is composed of multiple connected components, we also seep on each connected component in parallel.

%% file: tex/Experiments.tex
We implement QOCE with a mixture of Matlab 2017a and C++\footnote{https://github.com/PanShi2016/QOCE}.
We compare QOCE with four state-of-the-art community detection methods on a range of real-world networks from various domains and synthetic datasets with various parameters for performance evaluation. All experiments are conducted on a server with 2 Intel Xeon CPUs at 2.30GHz and 256GB main memory.

For the hyperparameters of QOCE, we fix the number of steps $t_0=3$ for random walk sampling, and the default value of the threshold $\mu$ is $0$ in Algorithm \ref{alg:random_walk}, such that the sampled subgraph is large enough to cover all possible members of the target community. We set $\alpha=0.2$ for the optimization problem in (\ref{Eq:final_opt}) and the window size $w=5$ for Algorithm \ref{alg:post_proc} to have a good trade-off on real-world networks as well as synthetic datasets.

\subsection{Synthetic Networks.}~
For synthetic datasets, we use the LFR benchmark networks \cite{lancichinetti2008benchmark}. The LFR benchmark graphs have a built-in binary community structure, which simulates the property of real-world networks on the heterogeneity of node degrees and community size distribution.

\begin{table}[htbp]
    \setlength{\abovecaptionskip}{0pt}
	\setlength{\belowcaptionskip}{5pt}
	\caption{Parameters for the LFR benchmark networks.}
	\label{tab:lfr}
	\centering
	\scalebox{0.8}
	{
	\begin{tabular}{l|l}
	\hline
	Parameter & Description\\
	\hline
	$n=5,000$ & number of nodes in the network\\
	$\mu = 0.3$ & mixing parameter\\
	$\bar d = 10$ & average degree of the nodes\\
	$d_{max} = 50$ & maximum degree of the nodes \\
	$s:[10,50], b:[20,100]$ & range of the community size \\
	$\tau_1 = 2$ & node degree distribution exponent\\ 
	$\tau_2 = 1$ &community size distribution exponent \\ 
	$om \in \{2,3...,8\}$ &overlapping membership\\
	$on \in \{500, 2500\}$& number of overlapping nodes\\
	\hline
	\end{tabular}
	}
\end{table}

We adopt the standard LFR parameter settings used in the survey of overlapping community detection~\cite{xie2013overlapping} and generate four groups with a total of 28 LFR benchmark networks. Our parameter settings are summarized in Table \ref{tab:lfr}. The four groups of LFR networks are distinguished by the range of community size ($s$ or $b$) and the number of overlapping nodes ($on=500$ or $2500$). For each group, we vary the overlapping membership $om$ from 2 to 8 to generate seven networks. Denote these groups by $\text{LFR}\_\{s,b\}\_on/n$ and networks by $\text{LFR}\_\{s,b\}\_on/n\_om$, such as $\text{LFR}\_s\_0.5$ for the group $\{s:[10,50],~on=2500,~om=\{2,3...,8\}\}$ and $\text{LFR}\_s\_0.5\_2$ for the network $\{s:[10,50],~on=2500,~om=2\}$.

\subsection{Real-world Networks.}~
We consider six real-world network datasets with labeled ground-truth in various domains. The TerroAttack and TerroristRel networks~\cite{zhao2006entity} are related to terrorist attacks and terrorist organizations, respectively. The Polbooks is a network of books about U.S. politics from the 2004 U.S. presidential election\footnote{http://orgnet.com} taken from Amazon.com. Polblogs is a network of hyperlinks between weblogs on U.S. politics, recorded in 2005 by Adamic and Glance \cite{adamic2005political}. Additionally, two well-known large networks called Amazon and DBLP are from the SNAP website\footnote{http://snap.stanford.edu/}. 

\begin{table}[htbp]
    \setlength{\abovecaptionskip}{0pt}
	\setlength{\belowcaptionskip}{5pt}
	\caption{Some statistics of real-world datasets.}
	\label{tab:real}
	\centering
	\scalebox{0.85}
	{
	\begin{threeparttable}
	\begin{tabular}{l | r r r r}
	\hline
	Network & \#Nodes & \#Edges & \#Comms  & $\overline{|C|}^{*}$ \\
	\hline
	TerrorAttack	& 51	& 1,178	& 2		& 24.000 \\
	TerroristRel	& 665	& 6552	& 21	& 30.619 \\
	Polbooks		& 105	& 441	& 4		& 26.000 \\
	polblogs		& 1222	& 16714	& 3		&398.333 \\
	Amazon		& 334,863	& 925,872	& 75,149	& 30.000 \\
	DBLP		& 317,080	& 1,049,866	& 13,477	& 53.000 \\
	\hline
	\end{tabular}
	\begin{tablenotes}
		\footnotesize
		\item[*] $\overline{|C|}$ denotes the average size of communities in the network.
	\end{tablenotes}
	\end{threeparttable}
	}
\end{table}

As some of the labeled communities in these networks are disconnected, we preprocess all the datasets as follows: The datasets are first made undirected and unweighted, and the largest connected component (LCC) is extracted. For each dataset, we confine each labeled community to a subset of its members inside this LCC. For any disconnected community,  we separate into the connected components, and regard each such connected component with at least three nodes as a separate ground-truth community. Table \ref{tab:real} summarizes the statistics of the six real-world networks and their ground-truth communities after data cleaning process.

\begin{figure*}[htbp]
    \centering
    \subfigure[$\text{LFR}\_s\_0.1$]{
    \includegraphics[width=1.5in]{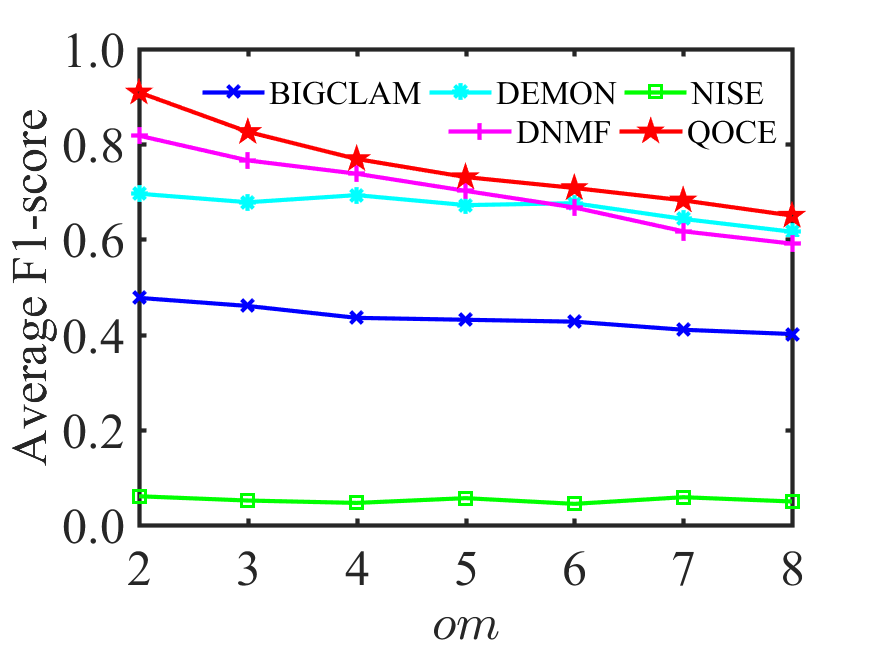}
    }
    \subfigure[$\text{LFR}\_s\_0.5$]{
    \includegraphics[width=1.5in]{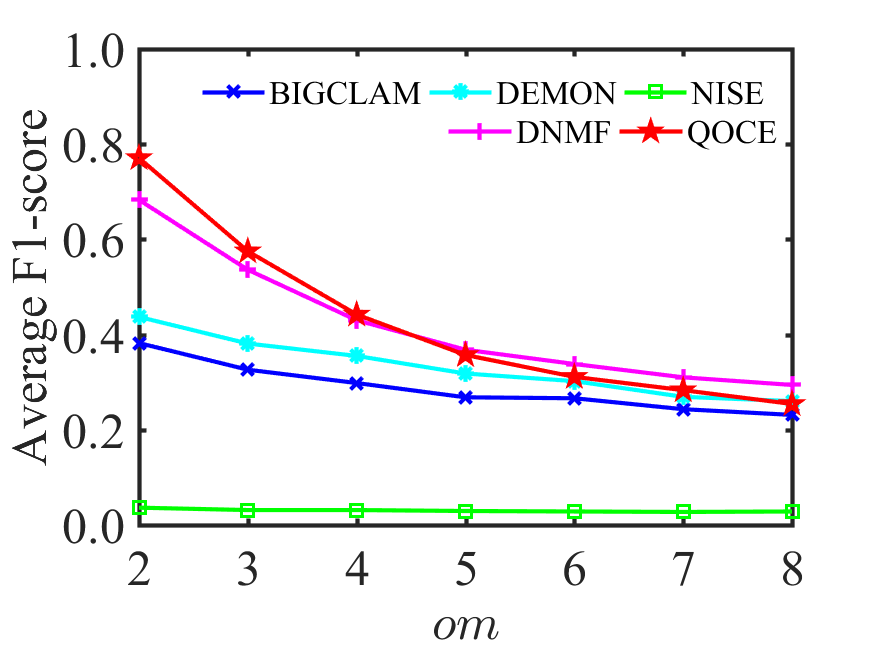}
    }
    \subfigure[$\text{LFR}\_b\_0.1$]{
    \includegraphics[width=1.5in]{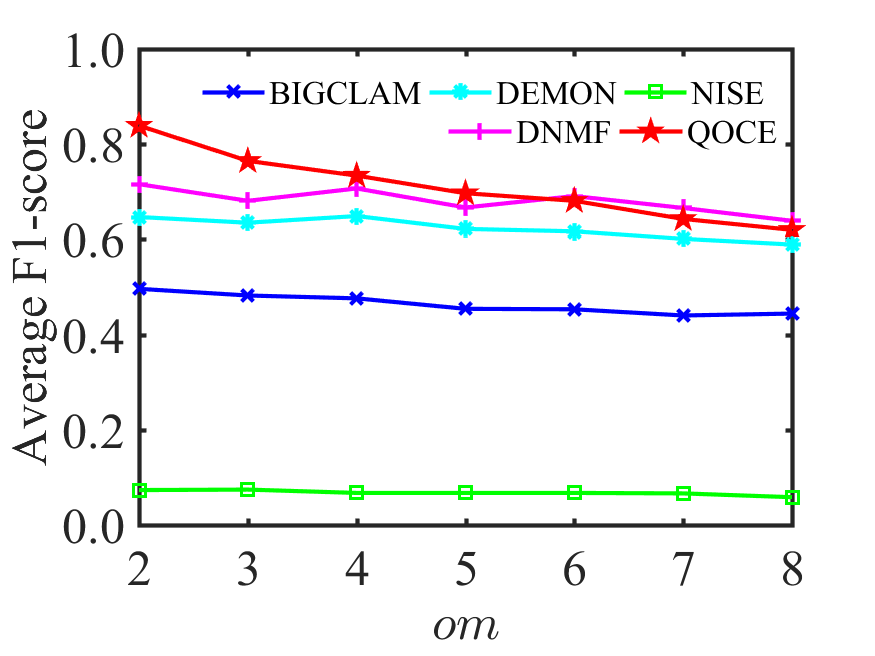}
    }
    \subfigure[$\text{LFR}\_b\_0.5$]{
    \includegraphics[width=1.5in]{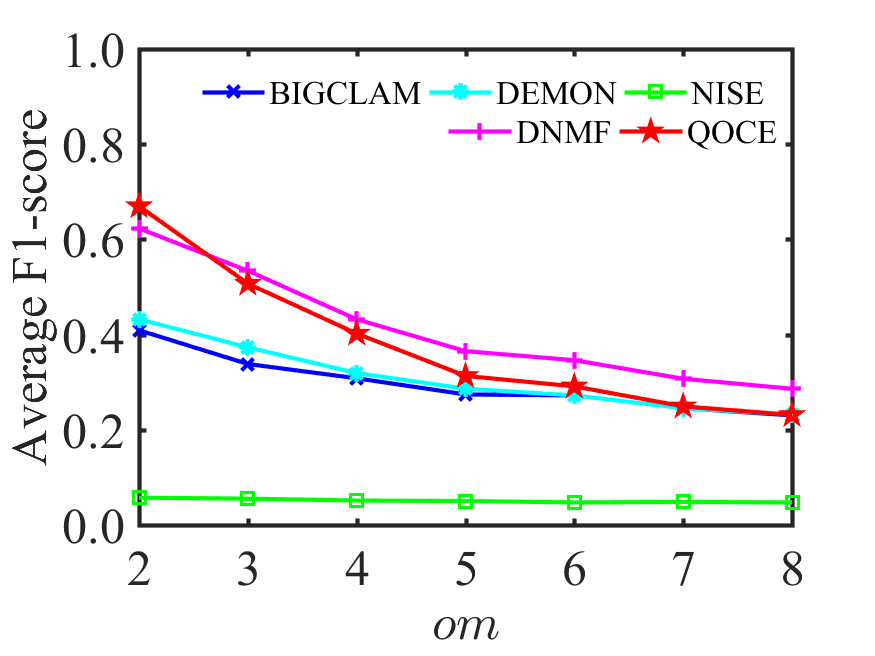}
    }
    \vspace{-1em}
    \caption{Comparison on the average F1-score on 28 LFR benchmark networks in four groups.}
    \label{Fig:lfr_f1}
    \vspace{-1em}
\end{figure*}

\subsection{Baseline Methods.}~
We select four state-of-the-art overlapping community detection algorithms as the baseline methods.

\begin{itemize}
\setlength{\itemsep}{0pt}
\setlength{\parsep}{0pt}
\setlength{\parskip}{0pt}
	\item \textbf{DEMON}~\cite{coscia2012demon}. DEMON is a local-first approach, which lets each vertex vote for the surrounded communities in its limited view of the global system using a label propagation algorithm, and then it merges the local communities into a global collection.
	\item \textbf{BIGCLAM}~\cite{yang2013overlapping}. BIGCLAM is a nonnegative matrix factorization based model, which builds on an observation that overlaps between communities are densely connected.
    \item \textbf{NISE}~\cite{whang2016overlapping}. NISE is a seed set expansion algorithm that develops a seeding strategy based on neighborhood-inflated for a personalized PageRank scheme that optimizes the conductance score of communities.
	\item \textbf{DNMF}~\cite{ye2019discrete}. DNMF is a nonnegative matrix factorization based model, but it seeks for a discrete community membership matrix for overlapping community detection directly.
\end{itemize}


\vspace{-1em}
\subsection{Evaluation Metric.}~
For evaluating the performance of the community detection algorithms on labeled datasets, we consider metrics that quantify the similarity between the detected communities and the ground-truth communities. Given a set of detected communities $\mathcal{C}=\{C_1, C_2, ..., C_{|\mathcal{C}|}\}$ and a set of ground-truth communities $\mathcal{C}^*=\{C_1^*, C_2^*, ..., C_{|\mathcal{C}^*|}^*\}$, we consider the following metric to measure the similarity between $\mathcal{C}$ and $\mathcal{C}^*$. 

\textbf{Average F1-score}~\cite{yang2013overlapping}.~
For each detected community $C_i \in \mathcal{C}$, find the most similar ground-truth community and quantify the similarity by F1-score. Then, for each ground-truth community $C_i^* \in \mathcal{C}^*$, also find the most similar detected community and compute the F1-score. Formally, the average F1-score is defined as:
\begin{equation}\nonumber
\frac{1}{2|\mathcal{C}^*|} \sum_{C_i^* \in \mathcal{C}^*} \max_{C_j\in \mathcal{C}} F_1(C_i^*, C_j) + \frac{1}{2|\mathcal{C}|} \sum_{C_i \in \mathcal{C}} \max_{C_j^* \in \mathcal{C}^*} F_1(C_i,C_j^*),
\end{equation}
where the F1-score between two communities is:
$$F_1(C_i, C_j) = \frac{2|C_i \cap C_j|}{C_i| + |C_j|}.$$


\subsection{Accuracy Comparison on LFR Benchmark Networks.}~
For the four competitive baselines, we use codes provided by the authors. Note that the number of seeds $k$ in NISE and the number of communities $k$ in BIGCLAM and DNMF are hyperparameters. We set $k$ to 200, which is also the number of communities for most ground-truth for all the LFR benchmark networks. Other hyperparameters are set as default values according to their guidelines. 

Figure \ref{Fig:lfr_f1} shows the accuracy evaluated by the average F1-score of our QOCE method and the baselines on the 28 LFR benchmarks in four groups. Regarding the average F1-score, QOCE outperforms BIGCLAM, DEMON, and NISE on all the LFR benchmark networks. As for the comparison with DNMF, in general, the average accuracy of QOCE is slightly better than that of DNMF. The average F1-score of QOCE is higher than DNMF on 15 networks and lower than DNMF on 13 networks. 
DNMF works slightly better on relatively highly overlapping networks, and QOCE works better on somewhat less overlapping networks. Nevertheless, we will see in the next subsection that DNMF can not scale to large real-world networks.

QOCE detects communities through seed set expansion, and each clique seed set generates one community. When a node belongs to more communities (i.e., the $om$ is larger), it usually has more edges, which means the node would belong to a particular maximal clique and becomes a clique seed set. Thus, the number of detected communities to which a highly overlapping node belongs is usually smaller than that of ground-truth communities to which the node belongs. This phenomenon may explain why our method's accuracy decreases faster than DNMF when $om$ increases.


\begin{figure*}
    \centering
    \subfigure[Joint effects of $t_0$ and $\mu$]{
    \includegraphics[width=2.4in]{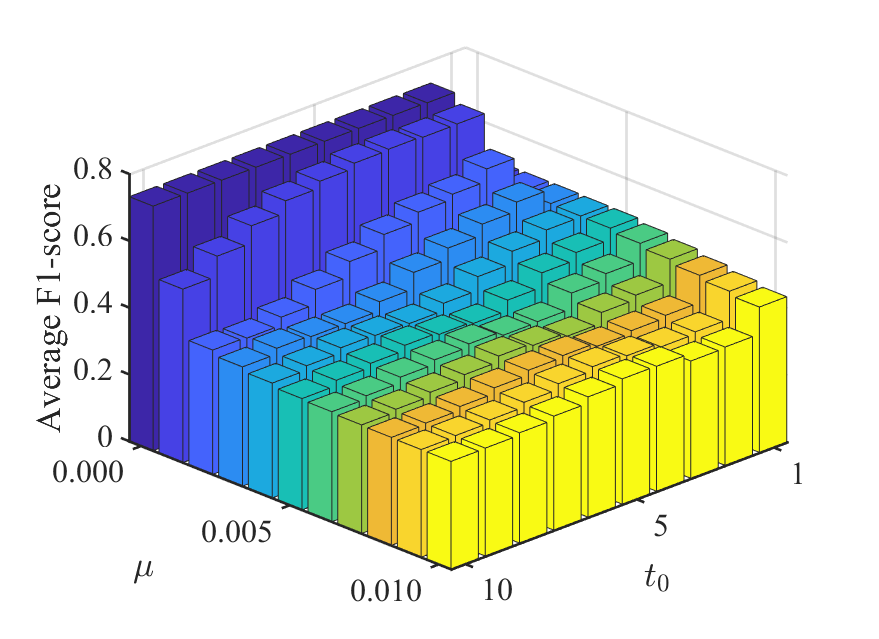}
    }
    \subfigure[Effects of $\alpha$]{
    \includegraphics[width=1.8in]{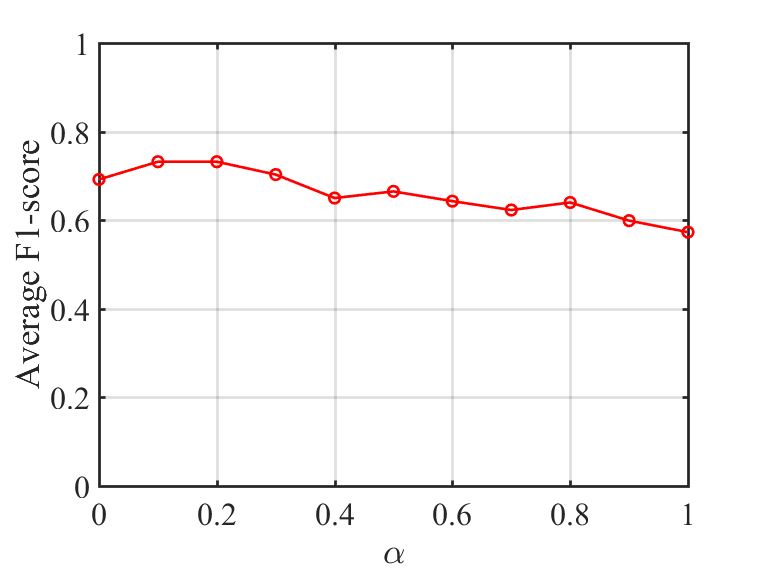}
    }
    \subfigure[Effects of $w$]{
    \includegraphics[width=1.8in]{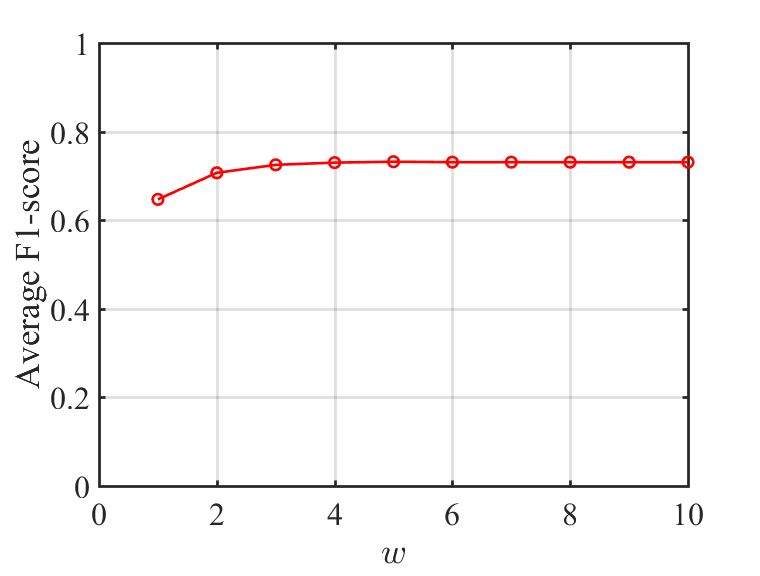}
    }
    \vspace{-1em}
    \caption{Sensitive study of the parameters.}
    \label{Fig:parameters}
    \vspace{-1em}
\end{figure*}

\subsection{Accuracy Comparison on Real-world Networks.}~
We set the key parameter $k$ of some baselines in different values favorable to these baselines for comparisons on real-world data. For small and medium-scale networks such as TerrorAttack, TerroristRel, polbooks, and polblogs, since the number of ground-truth communities is small, the hyperparameter $k$ in BIGCLAM is set as $-1$ (detect automatically), while the $k$ in NISE and DNMF (which do not support automatic detection) is set as 10. For large networks such as Amazon and DBLP, as in \cite{whang2016overlapping}, we set parameter $k$ in BIGCLAM, NISE, and DNMF as 25,000. Other hyperparameters are still set by their default values. 

\begin{table}[htbp]
    \setlength{\abovecaptionskip}{0em}
	\setlength{\belowcaptionskip}{1em}
	\caption{Comparison on the average F1-score on real-world networks.}
	\label{tab:real_f1}
	\centering
	\scalebox{0.7}
	{
	\begin{threeparttable}
	\begin{tabular}{l | c c c c c}
	\hline
	\textbf{Network}	& \textbf{BIGCLAM}	& \textbf{DEMON}	& \textbf{NISE}	& \textbf{DNMF}	& \textbf{QOCE} \\
	\hline
	TerrorAttack	& 0.438 	& \textbf{0.635}	& 0.429	& 0.346	& \underline{0.615} \\
	TerroristRel	& 0.453 	& \textbf{0.557}	& 0.462	& 0.417	& \underline{0.473} \\
	polbooks	& 0.470 	& 0.675	& \underline{0.714}	& 0.482	& \textbf{0.749} \\
	polblogs	& 0.321 	& 0.438	& \textbf{0.647}	& 0.333	& \underline{0.564} \\
	Amazon	& 0.423	& \underline{0.442}	& 0.338	& NaN*	& \textbf{0.473} \\
	DBLP	& \underline{0.355}	& 0.242	& 0.242	& NaN*	& \textbf{0.379} \\
	\hline
	\end{tabular}
	\begin{tablenotes}
		\footnotesize
		\item[*] out of memory.
	\end{tablenotes}
	\end{threeparttable}
	}
	\vspace{-1em}
\end{table}


Table \ref{tab:real_f1} shows the accuracy evaluated by the average F1-score of QOCE and the baselines on six real-world networks. As Table \ref{tab:real_f1} shows, QOCE always performs the best or the second-best in terms of the average F1-score, and QOCE performs better on larger networks. Note that DNMF cannot scale to large networks due to its excessive space overhead. When running on the Amazon network, DNMF requires 835.5GB memory to load the adjacency matrix. 


\begin{figure}
    \centering
    \includegraphics[width=3.2in]{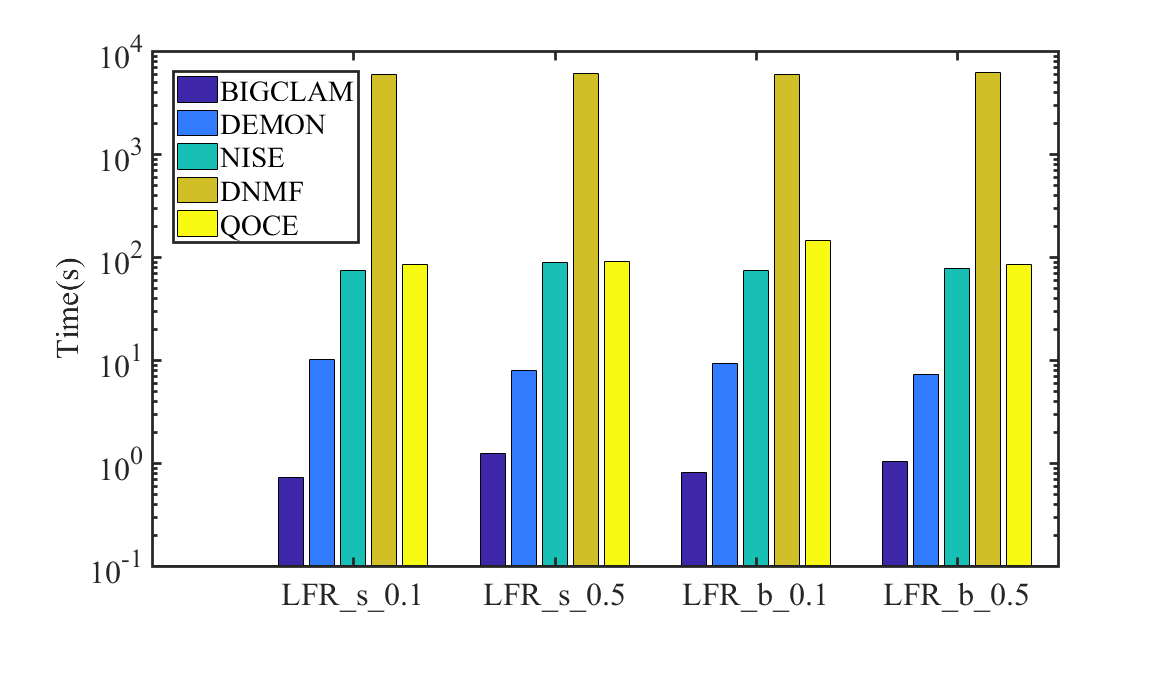}
    \vspace{-2em}
    \caption{The running time comparison on LFR benchmark networks. 
    Note that there is average running time on each of four groups of LFR networks. e.g., the group $\text{LFR}\_b\_0.5$ in the bar represents the average running time of different methods on the seven networks in Group $\text{LFR}\_b\_0.5$.
    }
    \label{Fig:time}
    \vspace{-1em}
\end{figure}

\subsection{Comparison on Running Time.}~
DEMON only supports sequential execution, BIGCLAM, NISE, and QOCE can be executed in parallel with multiple threads, and DNMF can be accelerated by the parallel matrix operation built in MATLAB. We set the number of threads for BIGCLAM, NISE, DNMF, QOCE to 20, and DEMON executes serially for all networks. Since these methods are implemented in different programming language, the comparison is for reference only. 

Figure \ref{Fig:time} illustrates the running time of different methods on LFR benchmark networks and real-world networks. On LFR benchmark networks, QOCE has a similar average running time with NISE, and runs much faster than DNMF. BIGCLAM, which is implemented in C++, has the shortest running time.
Furthermore, these methods show a similar trend on real-world networks, DNMF is always the slowest method and runs out of memory on Amazon and DBLP.

\subsection{Sensitive Analysis on Parameters.}~
We also analyze the parameter sensitivity of QOCE on $\text{LFR}\_s\_0.1\_5$. Similar results can be observed on other networks. There are a total of four hyperparameters in QOCE, including the random walk step $t_0$, the random walk threshold $\mu$, the trade-off parameter $\alpha$, and the window size $w$, as shown in Algorithm \ref{alg:qoce}.

The random walk step $t_0$ and threshold $\mu$ jointly control the size of the subgraph. We fixed $\alpha=0.2$ and $w=5$, and evaluate QOCE with the average F1-score by varying $t_0$ in [1,10] and $\mu$ in [0,0.01] simultaneously. The results are presented in Figure \ref{Fig:parameters} (a). We can see that QOCE performs poorly when $t_0=1$, because the subgraph obtained by one-step random walk sampling is too small to extract the community. Similarly, when $t_0$ and $\mu$ increase, the average probability of the vertices decreases, and the threshold is increased, which means the subgraph is too small to extract the proper community. Meanwhile, when $\mu$ takes 0 as the default setting, QOCE performs best and is robust to $t_0$.

The trade-off parameter $\alpha$ effects the value of affiliation vector $\mathbf{y}$. We obtain the effects of $\alpha$ by fixed $t_0=3, \mu=0, w=5$ and varying $\alpha$ in [0,1], and present it in Figure \ref{Fig:parameters} (b). We can observe that the performance of QOCE decreases slightly with the increase of $\alpha$, and the decline is less than 0.16. When $\alpha$ takes 0.1 or 0.2 (same as default), QOCE shows the best performance. 

The window size $w$ controls the size of the detected communities. We fixed $t_0=3, \mu=0, \alpha=0.2$ and varying $\alpha$ in [1,10]. The effects of $w$ are illustrated in Figure \ref{Fig:parameters} (c). Obviously, QOCE is not sensitive to $w$, especially when $w \geq 3$, the average F1-score hardly changes and reaches the optimal.

%% file: tex/Conclusion.tex
In this paper, we developed a new seed set expansion method called Quadratic Optimization based Clique Expansion (QOCE) for the overlapping community detection problem. QOCE is in an unsupervised manner and does not require any prior knowledge of the number of communities. Due to maximal cliques' dense internal connections, we use high-quality maximal cliques as the seed sets. In contrast to other expansion methods, which expand each seed set globally and greedily, our QOCE method first samples a subgraph from each seed set. This sampling reduces the time and space overhead of the follow-up expansion. The QOCE method then extracts a community from the sampled subgraph by solving a quadratic optimization problem. To maximize this benefit, we also parallelize our method by processing several maximal clique seeds independently and simultaneously. 

Extensive experiments show that QOCE yields accurate results on community detection in real-world networks of various domains and scales, and diverse synthetic networks with systematically configured parameters and QOCE is very competitive compared with the state-of-the-art baselines. Note that for the baselines, we set their key parameter of the number of communities in different values based on the prior of different datasets.